\documentclass[useAMS,usenatbib,usegraphicx,newapa,letterpaper]{mn2e}

\usepackage{graphicx}
\usepackage{color}
\usepackage{soul}
\usepackage{colortbl}
\definecolor{LightCyan}{rgb}{0.78,1,1}
\definecolor{LightMagenta}{rgb}{1,0.78,1}

\usepackage{multirow}

\usepackage{natbib}

\title[Time series photometry of EC\,04207]{Time series photometry of the 
       helium atmosphere pulsating white dwarf EC\,04207$-$4748}
\author[P.~Chote et al.]
       {P.~Chote,$^{1}$\thanks{Email: paul.chote@vuw.ac.nz; 
visiting astronomer, Mt John University Observatory (MJUO), operated by the 
Department of Physics \& Astronomy, University of Canterbury.}
        D.~J.~Sullivan,$^{1}$\thanks{Email: denis.sullivan@vuw.ac.nz; 
visiting astronomer, MJUO}
        M.~H.~Montgomery,$^{2}$
        and J.~L.~Provencal$^{3}$  \\
$^1$School of Chemical \& Physical Sciences, Victoria University of Wellington,
    P.O. Box 600, Wellington, New Zealand.\\
$^2$Department of Astronomy and McDonald Observatory, University of Texas,
    Austin TX 78712, USA. \\
$^3$University of Delaware, Department of Physics and Astronomy, Newark
    DE 19716, USA.}
    
\begin{document}

\date{Accepted 2013 January 29}

\pagerange{\pageref{firstpage}--\pageref{lastpage}} \pubyear{2012}
\maketitle

\label{firstpage}

\begin{abstract}
We present the analysis of 71 hours of high quality time-series CCD
photometry of the helium atmosphere pulsating white dwarf (DBV)
EC\,04207$-$4748 obtained using the facilities at Mt John University
Observatory in New Zealand.  The photometric data set consists of four
week-long observing sessions covering the period March to November
2011.  A Fourier analysis of the lightcurves yielded clear evidence of
four independent eigenmodes in the star with the dominant mode having
a period of 447\,s.  The lightcurve variations exhibit distinct
nonsinusoidal shapes, which results in significant harmonics of the
dominant frequency appearing in the Fourier transforms.  These
observed variations are interpreted in terms of nonlinear
contributions from the energy flux transmission through the subsurface
convection zone in the star.  Our modelling of this mechanism, using
the methods first introduced by \cite{montgomery05}, yields a
time-averaged convective response time of $\tau_{0} \sim 150$\,s for
the star, and this is shown to be broadly consistent with an
MLT/$\alpha$ parameter value between 0.8 and 1.2. It is argued that
for the DBV pulsators the measured value of $\tau_{0}$ is a better
estimate of the relative stellar surface temperatures than those
obtained via spectroscopic techniques.
\end{abstract}

\begin{keywords}
  asteroseismology -- convection -- instrumentation: photometers -
  stars: individual: EC\,04207$-$4748 -- white dwarfs.
\end{keywords}

\section{Introduction}

Stellar evolutionary models indicate that something approaching
98\% of all stars will end their active lives as slowly cooling white
dwarfs. This makes them particularly interesting astrophysical objects.
The subset of white dwarfs that pulsate are even of more
interest as the detected light flux variations can be used to identify
stellar normal modes which can then be matched to the spectrum of
possible modes predicted by theoretical models of the star.  In this
way the models can be constrained and improved, thus allowing us to
explore stellar structure below the photosphere.  This process is
referred to as asteroseismology.  Furthermore, as the difference
between a pulsator and a non-pulsator in each class appears to be
simply a matter of where it appears on the white dwarf cooling curve,
what we learn from the pulsators should be applicable to all white
dwarfs in that class.  Good general reviews on the topic of white
dwarfs, including the pulsators, can be found in \cite{winget08},
\cite{fontaine08} and \cite{althaus10}.

Currently there appear to be \emph{four} classes of pulsating white
dwarf stars.  At the hot end of the spectrum there are the GW\,Vir (or
PG\,1159) stars with $T_{\rm eff} \sim 120$\,kK and at the cool end
there are the ZZ\,Ceti objects with T$_{\rm eff} \sim 12$\,kK.  In
between these two effective temperature extremes lie the V777
pulsators with T$_{\rm eff} \sim 25$\,kK and the recently discovered
hot DQV objects with carbon rich atmospheres and T$_{\rm eff} \sim
20$\,kK.  Strictly speaking the very hot GW\,Vir or DOV objects
are really pre-white dwarfs, and there appears to be some debate about
the origin of the variability in the DQVs.

The common physical property that explains the pulsation phenomenon in
these white dwarfs, and in particular the effective temperature
regimes where it occurs, is a zone of a partially ionized chemical
element just below the stellar surface.  For the ZZ\,Ceti or DAV
objects it is the sub-surface H zone that is responsible for
pulsation, while it is a partially ionized He zone in the case of
the V777 or DBV stars.

The first DAV pulsator (HL\,Tau\,76) was serendipitously discovered in
1968 \citep{landolt68}, and in an impressive confirmation of the
partial ionization zone requirement, the first DBV pulsator was
discovered after a targeted search of known helium atmosphere (DB)
white dwarfs \citep{winget82}.  Note that for the DOVs it is partial
ionization of the heavier elements carbon and oxygen that is the key,
and for the DQVs, with carbon rich atmospheres, an ionization state of
C is predicted to drive the pulsations.

It is interesting to also note that the family of white dwarf pulsators
has only recently been extended to include an extremely low mass (ELM)
white dwarf \citep{hermes12}.  These very low mass objects have a
helium core and a hydrogen envelope so a sub-surface hydrogen partial
ionization zone presumably drives the pulsation.  Given the
significant difference between an ELM helium core white dwarf and the
much more common DA white dwarf with a carbon/oxygen core, it is debatable
as to whether a new instability strip has been discovered or the 
existing DAV strip has been extended in some way.

Typical white dwarf pulsators are multi-periodic with pulsations that
correspond to buoyancy-driven non-radial gravity modes. These pulsations
have periods in the $\sim10^{2}$ to $\sim10^{3}$ second range.

The DAVs dominate the population of known pulsators as there are now
almost 150 that have been discovered
\citep[e.g.][]{mukadam04a,mullally06}, with a large number detected in
the last decade following the identification of many faint white
dwarfs in the Sloan Digital Sky Survey (SDSS).  This large number of
DAVs has enabled meaningful investigations of the nature of the narrow
DAV instability strip in the temperature regime of T$_{\rm eff} \sim
12$\,kK for the hydrogen atmosphere white dwarfs.

In contrast, there are only 21 currently known DBVs
\citep[e.g.][]{nitta09,kilkenny09}, with the most recent object
discovered in one of the Kepler fields \citep{ostensen11}.
Consequently, the white dwarf atmospheric parameters characterising
the DBV instability strip in the temperature region of T$_{\rm eff}
\sim 25$\,kK for the helium atmosphere objects is not very well
defined.  It is therefore important to investigate the properties of
as many of these stars as possible.

In this paper we present approximately 71 hours of new CCD time-series
photometry obtained on the pulsating DBV white dwarf EC\,04207--4748
(simply EC\,04207 subsequently).  We also provide an analysis of the
lightcurves yielding physical properties of the star.

\section[]{Observations}

\subsection[]{Previous Work}

The Edinburgh-Cape southern hemisphere survey \citep{stobie97}
identified EC\,04207 as a 15.3 V magnitude helium atmosphere
white dwarf from a low-dispersion spectrogram obtained during the
survey \citep{kilkenny09}.  The spectrum featured only broad He I
absorption lines, which is a key spectral signature for DB white
dwarfs in the vicinity of the instability strip.

The Hamburg European Southern Observatory survey also independently
identified this object as a helium atmosphere white dwarf (which they
named HE\,0420--4748) and two separate spectral analyses have been
performed yielding the following atmospheric parameters: (1) $T_{\rm
  eff} \sim 25$\,kK and $\log g = 8.2$ \citep{koester01}, and (2)
$T_{\rm eff} \sim 27.3$\,kK and $\log g = 7.8$ \citep{voss07}.  This
spread of atmospheric parameter estimates from the analysis of DB
white dwarf spectra is fairly typical.

These effective temperature values for EC\,04207 place this
white dwarf within the (admittedly not that well defined) DBV
instability strip, so it is unsurprising that \cite{kilkenny09}
actually identified it as a pulsator via 10 hours of time-series
photometry obtained in 2002/2003.  A Fourier analysis of this
photometry (consisting of 3 separate runs) revealed a dominant
frequency of $\sim$2235\,$\mu$Hz, an apparent harmonic at twice this
value and a suggestions of a few other periodicities.  It is clear
that a more substantial observing programme on this object should lead
to an improved characterisation of its lightcurve frequency
structure.

\subsection[]{New Mt John CCD Photometry} \label{sec:mtjphot}

CCD time-series photometry of EC\,04207 was obtained during four
separate observing sessions in 2011 using the Puoko-nui photometer
attached to the one metre McLellan telescope at Mt John University
Observatory (MJUO, operated by the University of
Canterbury).  The Puoko-nui instrument is briefly described in Section
\ref{sec:puokonui}.  We obtained 71 hours (only 56 if the useful
exposure intervals are simply aggregated) of quality photometry
distributed over the four observing sessions.

Weather conditions during the first (March 2011) run were average with
poor seeing, while the remaining three runs (two in July 2011 and one
in November 2011) featured good observing conditions.  All data were
acquired with an exposure time of 20\,s and $2\times2$ CCD pixel
binning (see Section \ref{sec:puokonui}).  A summary of our
data set is provided in Table \ref{table:obs}, which
divides the observations into individual runs, the
UT date of the observations, the UT start time of the first
observation in each run, and the duration and number of 20\,s CCD frames 
obtained in each run.

The CCD exposures were processed in the standard way, involving bias,
dark current and flat field corrections.  Since the field is quite
sparse we extracted photometric data for the target and two
comparison stars from the science frames using the synthetic aperture
method.  Although this procedure is relatively standard, we briefly
summarise our techniques in Section \ref{sec:software}.

The reduced photometry files for each individual run were aggregated
into a combined file corresponding to each of the week-long observing
campaigns.  The resulting four files are summarised in Table
\ref{table:ts}, which includes the exposure start times in barycentric
Julian days (bjd), the number of exposures and effective total
observation times, and the duty cycle for each of the four campaigns.

\begin{table}
\label{table:obs}
\caption{A summary of the observations obtained on the pulsating white
  dwarf EC\,04207.  All CCD frames were acquired with 20\,s exposures
  and the individual observing sequences are listed with run name, UT
  date, UT start time of the first exposure in each run, the length of
  the run in hours and the number of exposures in each run.}
\centering
\begin{tabular}{lrrrr}
\hline
\multicolumn{1}{c}{Run Name} & 
\multicolumn{1}{c}{UT Date} & 
\multicolumn{1}{c}{UT Start} & 
\multicolumn{1}{c}{$\Delta$T} & 
\multicolumn{1}{c}{N} \\
          & (2011)  &     & [h] &   \\
\hline
20110302  & 2 Mar &  9:00:50 & 2.98 & 496 \\
20110306  & 6 Mar &  8:48:40 & 2.09 & 256 \\
20110307  & 7 Mar &  8:49:30 & 3.27 & 519 \\
20110308  & 8 Mar &  8:09:30 & 3.89 & 701 \\
\hline
20110702  & 2 Jul & 16:16:20 & 2.94 & 524 \\
20110703  & 3 Jul & 15:50:40 & 3.47 & 625 \\
20110704  & 4 Jul & 15:40:30 & 3.52 & 596 \\
20110705  & 5 Jul & 15:57:48 & 0.74 & 135 \\
20110707  & 7 Jul & 15:41:40 & 3.44 & 617 \\
\hline
20110727   & 27 Jul & 13:37:10 & 5.40 & 860 \\
20110728a  & 28 Jul & 13:04:30 & 1.01 & 176 \\
20110728b  & 28 Jul & 17:59:50 & 0.88 & 160 \\
20110729   & 29 Jul & 13:58:40 & 0.84 & 152 \\
20110730   & 30 Jul & 13:16:30 & 5.54 & 931 \\
20110731   & 31 Jul & 16:42:10 & 1.69 &  69 \\
20110801   &  1 Aug & 13:19:00 & 4.49 & 650 \\
20110802   &  2 Aug & 14:18:47 & 4.71 & 715 \\
\hline
20111118a  & 18 Nov & 10:02:20 & 2.90 & 389 \\
20111118c  & 18 Nov & 15:06:40 & 1.02 & 168 \\
20111121   & 21 Nov &  9:10:20 & 6.29 & 1080 \\
20111123a  & 23 Nov &  9:13:20 & 1.81 & 326 \\
20111123b  & 23 Nov & 11:40:20 & 1.49 & 270 \\
20111124   & 24 Nov &  9:28:20 & 6.34 & 1140 \\
\hline
\end{tabular}
\end{table}


\begin{table}
\label{table:ts}
\caption{Files containing the combined photometry for each of the four
  one week observing campaigns. Column 1 gives the file name, 2 is the
  start time of the first exposure in barycentric Julian day (BJD)
  units, 3 and 4 provide the number of useful exposures and effective
  observation times, while the last column states the duty cycle for
  each of the individual campaigns.}
\centering
\begin{tabular}{llrrr}
\hline
\multicolumn{1}{c}{File Name} & 
\multicolumn{1}{c}{Start Time} & 
\multicolumn{1}{c}{N} &
\multicolumn{1}{c}{$\Delta$T} &
\multicolumn{1}{c}{Duty} \\

\multicolumn{1}{c}{} & 
\multicolumn{1}{c}{[BJD]} & 
\multicolumn{1}{c}{} &
\multicolumn{1}{c}{[hr]} &
\multicolumn{1}{c}{Cycle} \\
\hline
mar11.ts  & 2455622.8752867 & 1972 & 11.0 &  8\% \\
jul11a.ts & 2455744.2439362 & 2497 & 13.9 & 10\% \\
jul11b.ts & 2455770.0678829 & 3713 & 26.6 & 18\% \\
nov11.ts  & 2455883.9212131 & 3373 & 18.7 & 13\% \\
\hline
\end{tabular}
\end{table}


\section{The puoko-nui photometer}\label{sec:puokonui}

\subsection{Hardware \& Firmware}\label{sec:hardware}

The core of the Puoko-nui (meaning \emph{big eye} in Maori) photometer
is a Princeton Instruments (PI) Micromax $1024\times1024$ pixel frame
transfer CCD.  Frame transfer CCDs operate by rapidly shifting
accumulated charge into a masked region of the CCD, allowing
digitization to operate in parallel with exposure of the next
frame. This allows for a shutterless design with negligible
deadtime. The initial plan was to develop a CCD time-series photometer
largely for use on the McLellan one metre telescope at MJUO in NZ that
was based on the Argos instrument \citep{nather04}.  However, although
the underlying concept of using an external GPS-conditioned signal to
trigger each successive frame transfer has been retained, many of the
implementation details are different.  As they have not been described
elsewhere it is appropriate to briefly summarise them here.

The photometer is attached at the Cassegrain f/8 focus position of the
MJIO one metre telescope via an offset-guider box that was part of a
previous instrument -- the VUW three-channel photometer
\citep{sullivan00a}.  At this focal position the plate scale is 25.8
arcsec per mm, so the PI camera's 13 micron square pixels each image a
0.33 arcsec square segment of the sky, which leads to an overall field
of view of 5.7\,arcmin square for the CCD.  Given that the typical
seeing conditions at MJUO are $\sim 2$ arcsec, we typically operate
the camera in $2\times2$ pixel binning mode with 0.66 arcsec per pixel,
and these frames can be downloaded in 3.2\,s at the slowest
(100\,kHz) readout rate via USB.

The shortest periods in our usual pulsating white dwarf targets are
$\sim 100$\,s, so we normally employ exposure times no smaller than
10\,s.  However, the camera can be operated at much shorter exposure
times than this without introducing significant deadtimes by using
the higher available digitization rate (1 Mhz) in combination with
further binning and/or reading out subsections of the CCD pixels.

As the hot white dwarfs produce relatively little flux in the red
wavelength region, we filter the light to the CCD with a broad band
blue BG40 glass filter in order to reduce the background introduced by
sky photons.

Use of the offset guiding box to mount the instrument allows use of a
wide-angle eyepiece to resolve any start-up issues that may arise.
However, more importantly, the configuration facilitates use of a
separate CCD camera with access to an expanded field of view, thus
allowing identification of a suitable bright star for telescope
autoguiding purposes separate from the main CCD.  Specifically, a
45$^\circ$ mirror with a central hole redirects an annulus around the
primary light path to an SBIG ST-402ME CCD mounted on a
two-dimensional slide mechanism.  The CCD can be manually aligned on a
nearby bright star, and the CCD-OPS software used to process acquired
frames and return guide star offset information.  We use this feature
to autoguide the MJUO one metre telescope as there is no facility
feature providing this capability.

The external frame transfer trigger signal for the camera is generated
by a custom timer unit built around an AVR ATMega 1284 microprocessor.
This device has inputs for a 1\,Hz pulse train and RS232 serial data
stream from an external GPS receiver, and it communicates with the
acquisition PC via USB.  The primary operating mode of the timer uses
the incoming pulse train to generate accurate UT-aligned frame
transfer pulses at integer-second intervals, and time information
extracted from the serial data stream is sent to the acquisition PC to
be matched with each acquired frame.  A further function of the timer
unit is to monitor the status of the camera via a provided hardware
output, as this information is not made available in the PI proprietary
software interface for the Micromax system.

Code for the microprocessor has been developed using the C language,
and can be cross-compiled and downloaded to the timer unit via its USB
connection using standard software tools.  Currently the code supports
the signals from the two GPS receivers we currently use -- a portable
Trimble Thunderbolt unit, and an older fixed installation Magellan OEM
receiver.  We also have developed a variant of our standard timer
configuration that extends the exposure intervals into the sub-second
domain.  The timer unit's processor is driven with a UT-synchronised
clock signal generated by one of the GPS units to ensure stable cycles
and correct UT alignment, and obtains a timing resolution of 1\,msec
with a precision better than 5\,$\mu$sec.  This function is useful for
operating the camera at higher than normal duty cycles and is
particularly relevant for later generation cameras such as the ProEM
system marketed by Princeton Instruments.

A key design feature of the photometer is the provision of separate
hardware for the timing and data acquisition/analysis functions.  This
modular design adds considerable flexibility, making it relatively
simple to support other CCDs or GPS receivers, and places no
restrictions on the control and acquisition PC (aside from those
required for the CCD, and having a free USB port).  In addition, the
latency timing delays resulting from use of the microprocessor
interrupt software mode are very small ($\sim\mu$sec) and consistent,
and are completely negligible for our purposes.

\subsection{Software}\label{sec:software}

The software in the acquisition PC has been developed in an Ubuntu
GNU/Linux environment (but can also run in the Windows or Mac OS
environments if appropriate camera drivers are available).  Raw CCD
frames are written to disk in a compressed FITS format, with
information such as timestamps and target star included in the file
header.  A separate program, \emph{tsreduce} processes each frame as
it is acquired and displays a real-time plot of the developing
lightcurve and its discrete Fourier transform.

\begin{figure*}
\centering
\includegraphics[width=12cm, angle=-90]{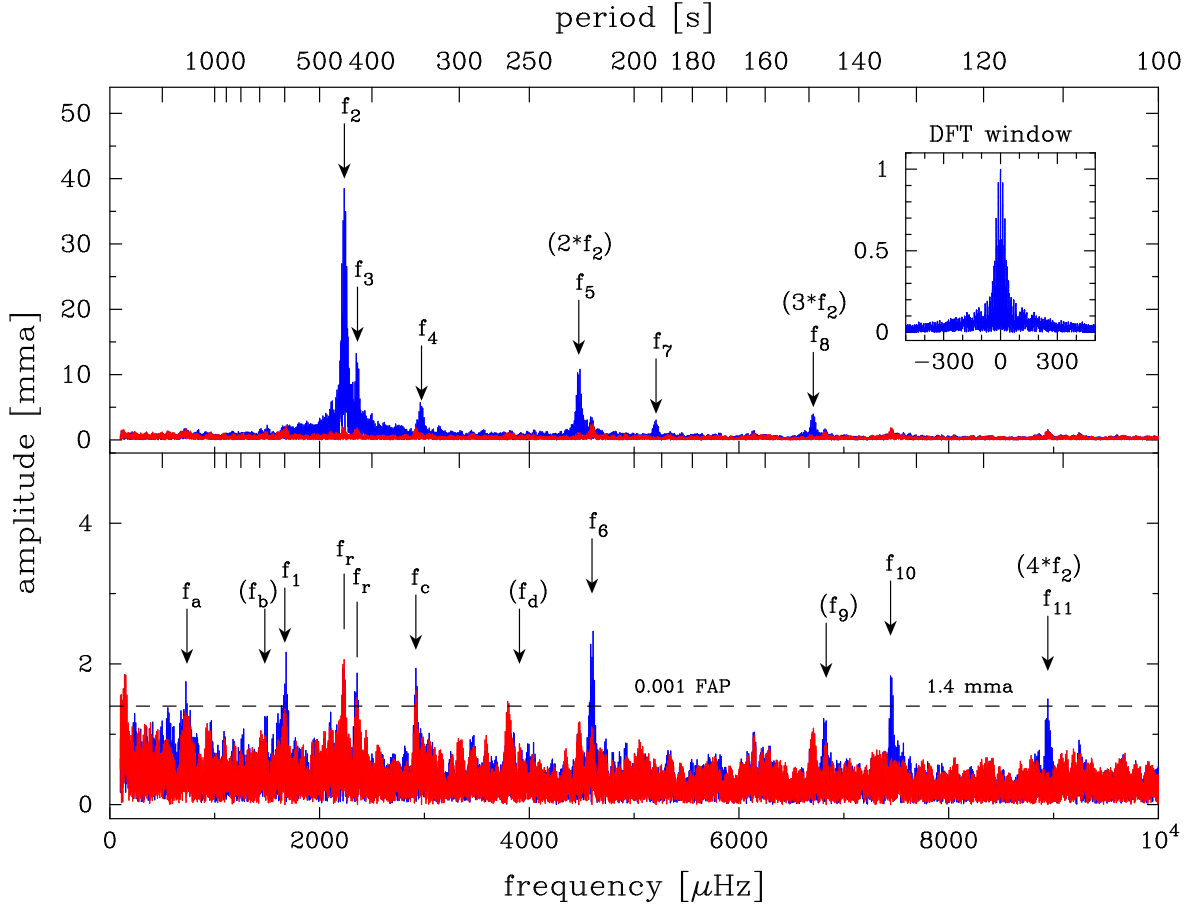}
\caption{Discrete Fourier transform of the jul11b.ts data set (Jul 27
  -- Aug 2, 2011). The top panel shows the unmodified DFT (or
  periodogram) in blue between $10^2 - 10^4\mu$\,Hz with the vertical
  scale in millimodulation amplitude (mma) units for which 10\,mma
  represents a 1\% modulation of the lightcurve.  The DFT after
  prewhitening by the 6 labelled frequencies is shown in red (lighter
  shade), and expanded vertically in blue (darker shade) in the bottom
  panel. The frequencies marked $f_{r}$ correspond to residual power
  which remained after subtracting $f_{2}$ and $f_{3}$ from the
  lightcurve and are not considered physical.  The four
  ``frequencies'' labelled $f_{a}$, $f_{b}$, $f_{c}$, $f_{d}$
  represent peaks seen in the DFTs in other runs, but we are reluctant
  at this stage to confidently identify these as representing real
  power arising from a pulsation mode.  The remaining power after
  prewhitening by all the identified peaks is again shown in red
  (lighter shade) in the bottom panel.  The frequencies $f_{5}$,
  $f_{8}$ and $f_{11}$ are clearly harmonics of the 38\,mma dominant
  frequency and have been appropriately co-labelled.  (A colour
  version of this figure is available in the online journal.)}
\label{fig:dft}
\end{figure*}

The initial rationale for the development of \emph{tsreduce} was for
the production of real-time lightcurves while observing was in
progress.  However, improvements made to the code mean that we now use
it as part of our final reduction pipeline for our sparse field
photometry, as comparisons with other available routines
have verified its accuracy.  We briefly outline the methodology here.

As mentioned in Section \ref{sec:mtjphot} the \emph{tsreduce} software
first processes each frame by subtracting a bias level (measured from
an overscan region in each frame), subtracting a mean dark frame to
remove dark current effects and finally dividing by an appropriate
flat field to correct for non-uniform illumination.  Synthetic
aperture photometry is then performed on the target and (typically
two) comparison stars using a central aperture of a fixed radius and a
surrounding annular region to estimate the background intensity.  The
aperture size for the online reduction is determined from the first
frame by identifying a minimum aperture radius such that the stellar
flux profile is indistinguishable from the background noise level.

The impact of atmospheric effects are minimised by normalising the
target light intensity with the sum of the two comparison star
intensities, and a low-order polynomial fit is then used to correct
for any residual effects such as differential extinction.  The
developing lightcurves of all three objects, along with the evolving
Fourier transform (DFT, see next section) of the target star, are
updated in real-time as new frames are acquired during an observing
run.  The FWHM of the seeing disk is also displayed on a plot as the
run proceeds.

Offline analysis routines allow the aperture size to be optimized for
the final reduction.  In this process we take the pragmatic approach:
the user is presented with the corrected lightcurve of the target
star, its DFT and the signal-to-noise estimates for different aperture
sizes.  Thus, an informed decision about the best aperture size can be
made.  We have yet to find an automated algorithm that can reliably
outperform the human eye and brain.

Also, if deemed appropriate, we use the program \emph{ts3fix}
\citep[described in][]{sullivan08a} in order to remove obvious
lightcurve artifacts such as resdiual transparency variations not
corrected for by the polynomial fitting.  Futher processing routines
convert the time stamps to barycentric Julian days (BJD) and combine
multiple runs as appropriate, allowing an analysis to be performed on
multiple nights of photometric data.


\section[]{Lightcurve Analysis}\label{sec:analysis}

\subsection{Fourier analysis}\label{sec:fourier}
\begin{figure*}
\centering
\includegraphics[width=5cm, angle=-90]{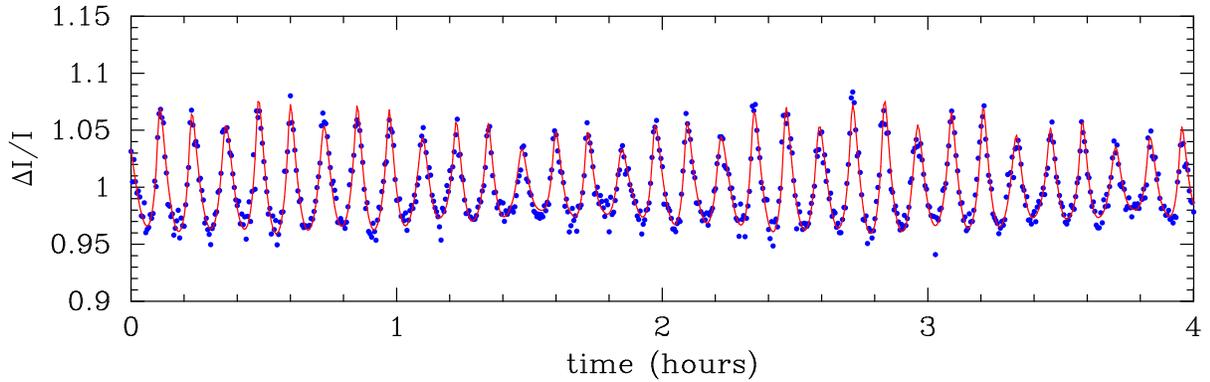}
\caption{Four hours of a EC\,04207 lightcurve that has been fitted by
  the convection modelling procedure discussed in the paper.  The
  vertical axis depicts the lightcurve modulation in fractional
  intentity units.  The non-sinusoidal pulse shapes in the lightcurve
  are clearly evident, and the fitting process yielded a value of
  $\tau_{0} = 148 \pm 5$\,s for the convective response time parameter
  (see text). (A colour version of this figure is available in the
    online journal.)}
\label{fig:lc}
\end{figure*}

Visual inspection of an EC\,04207 lightcurve (an example is given in
Fig.\ \ref{fig:lc}) clearly shows shows a non-sinusoidal oscillation
with a period of $\sim450$ seconds. It is thus no surprise that the
Fourier transform (shown in Fig.\ \ref{fig:dft} for the jul11b.ts run)
is dominated by a peak at $2236\,\mu$Hz and its harmonics.  Formally,
in Fig.\ \ref{fig:dft} we have displayed the amplitude power spectrum
in which the Fourier transform phase information has been removed; we
are only interested in the amplitudes of the sinusoids here.
Sometimes this plot is called a periodogram; we will simply call it a
DFT (short for discrete Fourier transform).  The horizontal plot scale
is $\mu$Hz and the vertical plot scale employs millimodulation
amplitude (mma) units, for which 10\,mma corresponds to a modulation
amplitude in the time domain of 1\% (which matches the amplitude of a
fitted sinusoid of the appropriate frequency and phase).

\begin{table*}
\label{table:freq}
\caption{Table of the detected frequencies in the four individual
  observing runs.  The amplitudes are specified in millimodulation
  amplitude (mma) units and those given in parentheses are below the
  0.1\% false alarm probability (FAP) for the particular run, but
  which we believe are likely to be real (see text).  The run FAP
  values are specified in the column headers in square brackets.
  Frequencies f$_{1}$ -- f$_{11}$ labelled by numbered suffixes are
  well determined across all four observing runs, whereas the four
  items labelled with letter suffixes a -- d exhibit significant power
  in at least one run, but we consider these only marginal detections
  of real power.  Frequencies deduced to be associated with
  independent eigenmodes of the star are labelled with a `$\star$'.}
\begin{tabular}{cccccccc}
\hline
 Item   & Frequency  & Period & \multicolumn{4}{c}{Amplitude [mma]} & Assumed \\
\cline{4-7}
        & [$\mu$Hz]  & [s]   & mar11.ts [2.7] & jul11a.ts [1.5] & jul11b.ts [1.4] & nov11.ts [1.4] & Combinations \\
\hline
f$_a$           &  736  & 1358.9  &  {-}  & {-}  &  1.7  & 1.6   &  {-} \\
f$_b$           & 1478  &  676.5  &  4.1  & {-}  & (1.3) & {-}   &  {-} \\
$\star$f$_{1}$  & 1669  & 599.1   &  4.7  & 2.6  & 1.9  & {-}   &      \\
$\star$f$_{2}$  & 2236  &  447.2  & 33.1  & 38.8 & 38.4 & 38.3  &  {-} \\
$\star$f$_{3}$  & 2361  & 423.5   & 8.7   & 8.1 &  8.1  & 9.2   &  {-} \\
f$_c$           & 2918   & 342.7  & (1.6) & {-}  &  1.7  & 2.1   &  {-} \\
$\star$f$_{4}$  & 2973  & 336.4   & 5.8   & 4.7 &  5.6  & 5.1   &  {-} \\
f$_d$           & 3906   & 256.0  & (2.1) & 1.6  &  {-}  & {-}   & f$_2$ + f$_1$   \\
f$_5$           & 4473   & 223.6  &  7.7  & 9.0  & 10.7  & 10.4  & 2\,f$_2$         \\
f$_6$           & 4598   & 217.5  & (2.1) & 3.2  &  2.6  &  2.0  & f$_2$ + f$_3$   \\
f$_7$           & 5209   & 192.0  &  2.7  & 2.1  &  3.1  &  2.5  & f$_2$ + f$_4$   \\
f$_8$           & 6709   & 149.1  &  2.9  & 2.8  &  4.1  &  4.2  & 3\,f$_2$         \\
f$_9$           & 6831   & 146.4  &  {-}  &(1.4) & (1.3) & 1.5   & 2\,f$_2$ + f$_3$ \\
f$_{10}$        & 7445   & 134.3  &  {-}  & {-}   &  1.7   & 1.4    & 2\,f$_2$ + f$_4$ \\
f$_{11}$        & 8945   & 111.8  &  {-}  & {-}   &  1.4   & 1.8   & 4\,f$_2$         \\
\hline
\end{tabular}
\end{table*}

The window function (Fig.\ \ref{fig:dft}, inset) shows the DFT of a
noise-free sinusoid sampled at the same times as the observed data.
It thus shows the `best' that one can do in visually identifying a
frequency from the Fourier transform.

We initially identified real frequencies in the jul11b.ts lightcurve
by visual inspection of the DFT.  Thus the six frequencies indicated
by downward arrows in the top panel of Fig.\ \ref{fig:dft} were
selected as real.  Given the non-sinusoidal shape of the lightcurve
it is clear that we expect to see harmonics of the dominant frequency
(f$_{2}$) in the DFT.  Hence f$_{5}$ and f$_{8}$ are clearly harmonics
of f$_{2}$ as their frequency values are exactly two and three times
that of f$_{2}$, respectively.  These are appropriately labelled in
Fig.\ \ref{fig:dft}.

Identification of real lower amplitude frequencies in a DFT is often
complicated by the aliases that arise from significant gaps in the
time domain data.  This applies here as we have combined multiple
nights in our single site observing runs -- the frequency resolution
is improved but prominent side peaks separated by $11.6\,\mu$Hz (1
day$^{-1}$) are also introduced due to the regular gaps with a period
of one day.  If there are real frequencies in the data with
separations comparable to the alias peaks, identification can be
difficult.

An obvious way to reduce the impact of alias peaks is to obtain more
observations during each 24 hour period.  This approach led to the
formation of the Whole Earth Telescope \citep[WET,][]{nather90}, which
is a collaboration of observers at multiple sites around the globe
with the aim of improving the 24 hour coverage of a particular target
over an entire multiple day observing run.  EC\,04207 was one of the
targets during the {\sc xcov}28 WET run in November 2011. We include
here results from the observations obtained at MJUO during {\sc
  xcov}28, primarily to show the variation in amplitudes of the modes
over our four observing runs.  Except for the disappearance and
appearance of several small amplitude frequencies, none of the results
presented here depend on the overall WET data set.

We have employed the technique of prewhitening to aid us in
identifying low level real frequencies in our data sets.  Six
sinusoids with frequencies corresponding to the large amplitude
signals discussed above were least squares fitted for both amplitude
and phase to the lightcurve.  The resulting model lightcurve was
then subtracted from the data.  Fixing the frequencies at the DFT
derived values allows direct use of the linear least squares
procedure.  The results can be further optimised by iterating the
frequencies using nonlinear least squares techniques, but in practice
this makes little difference.  Following this procedure, the signal and
aliases corresponding to the above frequencies will then have been
removed from the DFT of the prewhitened light curve.

This prewhitened DFT is plotted in red (lighter plot) in the top panel
of Fig.\ \ref{fig:dft} and is presented in the lower panel using a
different vertical scale.  The red (lighter) plot in the lower panel
represents a DFT further prewhitened by a number of the lower
amplitude frequencies.

Using this procedure, fifteen frequencies have been identified, and
are presented in Table \ref{table:convection} as well indicated by the
labelled downward arrows in Fig. \ref{fig:dft}.

For the low level signals it is important that noise peaks are not
mistaken as real in the DFTs.  We have used the false alarm
probability (FAP) method, first introduced by \cite{Scargle82}, to
quantify this issue.  With the speed of modern computers it is now
straight forward to employ Monte Carlo methods in this process rather
then assume some analytical noise model.  To determine the FAP level
for a given data set, the time series data is first prewhitened by all
of the frequencies with significant amplitudes.  This essentially
removes the majority of the coherent signal in the data.  The
timestamps for the resulting measurement values are randomly shuffled,
destroying any remaining coherent signal but leaving intact the
incoherent noise level.  The DFT of the time-shuffled data set is
determined and the maximum peak amplitude is recorded. Repeating this
exercise N times allows us to use the overall maximum peak value to
assert that there is a probablity of 1/N of obtaining a peak in the
DFT as high as this value that is due to incoherent noise.  It can
also be informative to plot a histogram of the collection of maximum
values, such as is presented in \cite{sullivan08a}.

We have chosen N to be 1000 and therefore can say that there is a
0.1\% chance of a noise ``conspiracy'' producing a DFT peak as high as
the calculated FAP value in each data set.  The calculated FAP value
(1.4 mma) for the jul11b.ts data set is represented by a dashed line
in the bottom panel of Fig.\ \ref{fig:dft}, and the mma values for all
four data sets are listed in square brackets in the headings in Table
\ref{table:freq}.

The frequencies we have identified as real at the 99.9\% confidence
level in the different data sets are listed in Table \ref{table:freq}
in terms of their mma values, and marked by downward arrows in
Fig.\ \ref{fig:dft} for the jul11b.ts data set.  Both in the table and
the figure we have used brackets to indicate frequencies that do not
strictly meet our FAP criterion but we have other reasons to believe
in their reality, such as they correspond to harmonic or sum frequency
values or they appear in another data set.

We note that subtracting the large amplitude frequencies, f$_2$ -
f$_8$, from the July and November runs in some cases left residual
power above the 0.1\% FAP threshold, particularly for the two largest
signals f$_2$ and f$_3$.  These  peaks are labelled by f$_r$ in
the bottom panel of Fig.\ \ref{fig:dft}.

This residual power was centred between 2 and 6\,$\mu$Hz from the
subtracted frequency, with several outliers that provided a best fit
as far as 50$\mu$Hz away for the DFT of the jul11a.ts data.  There does
not appear to be any consistent trend in the spacing of these residual
peaks between runs, or between frequencies within the same run, so we
are not prepared to associate them with physical stellar behaviour.
The power may result from instrumental or atmospheric effects in the
time domain that we have not properly modelled.  The combined data set
from the WET run may provide more insights.

The frequencies labelled with letter subscripts, namely $f_a$ --
$f_d$, are really in our uncertain category.  Their amplitudes appear in
subsets of the four runs above our adopted 0.1\% FAP threshhold, but
there are reasons to be cautious given their low amplitudes.  In
particular, $f_a$ corresponds to an unusually long period of 1360\,s
for this class of pulsator and is in the regime where 1/f noise may
have an impact on the data.  Even though it does appear above the FAP
threshhold in two data sets it is perhaps more likely to result from
transparency changes.

Taking into account the harmonic and combination frequencies and
setting aside the low amplitude `frequencies' $f_{a}$, $f_{b}$ and
$f_{c}$, it would appear that EC\,04207 is a relatively simple DB
pulsator with only four detected pulsation modes, $f_{1}$ -- $f_{4}$
(marked with a `$\star$' in Table \ref{table:freq}), and with a
lightcurve that is dominated by $f_{2 }$ (447\,s) and its harmonics.

\subsection{Nonlinear lightcurve fitting}\label{sec:nonlinear}

The Fourier analysis in the previous section clearly indicates that we
have only unambiguously identified four independent pulsation modes in
EC\,04207 from our four photometric data sets.  This is not really
enough detected eigenmodes of the star for us to place useful
constraints on its structure by matching these observed modes with
theoretically predicted values.  These detected modes could be
combined with those observed in other white dwarfs in the same DBV
class, and progress made by assuming they have a similar structure --
ie employ the techniques of \emph{ensemble asteroseismology}.  Instead
we will make use of an asteroseismic tool which focusses on the shape
of the observed lightcurve variations that was first introduced by
\cite{montgomery05}.

Building on work first undertaken by \cite{brickhill92a} and later by
\cite{goldreich99a} that emphasised the importance of the subsurface
convection zones in the white dwarf pulsators, \cite{montgomery05}
introduced a relatively simple model that enables the extraction of a
convection zone parameter from the variations in the lightcurve
shapes.  In essence Montgomery's technique models departures of the
lightcurve variations from a simple sinusoidal shape with effects
introduced by the energy flux transmission through the sub-surface
convection zone in the star.  The key derived parameter is a
time-averaged convective response time $\tau_{0}$, which corresponds
to an average time for energy to be trapped and then released by the
convection zone by way of the convection mechanism.  \cite{montgomery05}
first sought to model the lightcurves of mono-periodic (or nearly
mono-periodic) pulsators, but this was later extended
\citep{montgomery10} to include the multiperiodic ones.

The technique assumes that the radiative flux variations occurring at
the base of the convection zone are purely sinusoidal and that it is
the convective transport process itself that leads to the emergent
flux having a nonsinusoidal form.  Since the response time is expected
to increase with the total mass of the convective layer, the relative
times can then be used to differentiate between pulsators with
convection zones of various mass and therefore depth.  In simple
terms, the more nonsinusoidal (\emph{``nonlinear''}) the measured
lightcurve variations, the larger the value of $\tau_{0}$ and
therefore the more extended is the convection zone.

The EC\,04207 lightcurve variations are clearly non-sinusoidal, as is
evident from direct inspection of the lightcurve in
Fig.\ \ref{fig:lc} or by viewing the harmonics of the dominant
frequency in the DFT in Fig.\ \ref{fig:dft}.  The first step in the
convective lightcurve fitting is to select the \emph{independent}
pulsation modes to be incorporated; the fitting procedure itself
introduces harmonics and sum and/or difference frequencies as required
in order to match the observed lightcurve.  From the amplitudes of the
four detected independent pulsation modes displayed in the
Fig.\ \ref{fig:dft} DFT, it is clear that f$_{2}$ (and its harmonics)
along with f$_{3}$ and f$_{4}$ will explain essentially all of the
lightcurve variations.  Consequently, these frequencies were chosen
for the fit.  The parameters for the best fit are given in Table
\ref{table:convection}.

\begin{table}
\label{table:convection}
\caption{A list of the frequencies of the EC\,04207 pulsation modes
  used in the convection fitting procedure along with the derived
  parameters.  A key result of the lightcurve fitting is the value of
  $\tau_{0} = 148\pm5$\,s for the average convective response time for
  this star.}
\centering
\begin{tabular}{ccccccc}
\hline
Mode & Frequency & Period & Amplitude & $\ell$ & m \\
     & [$\mu$Hz] & [s]    & [mma]     &        &   \\  
\hline
 f$_{2}$ & 2236.24 & 447.18 & 23.8     & 1      & 0 \\
 f$_{3}$ & 2361.47 & 423.50 &  9.8     & 1      & 1 \\
 f$_{4}$ & 2972.66 & 336.40 &  4.0     & 1      & 0 \\
\hline
\end{tabular}
\end{table}

The solid line in Fig.\ \ref{fig:lc} shows the result of our fitting
procedure and it is clear that the model is a good fit to the observed
lightcurve and, in particular, it accurately reproduces the
non-sinusoidal shape.  Table \ref{table:convection} provides a list of
parameters describing the three eigenmodes for this best fit, and the
time-averaged convective response time for the fit is $\tau_{0} =
148\pm5$\,s.  Note that the value for $\tau_{0}$ in this fitting
procedure is not strongly dependent on the $\ell$ and m values for the
eigenmodes used in the fit.  Our fitting process also yielded an
estimate for the pulsation axis inclination angle of $\theta_i = 38.8
\pm 5$ degrees.

By itself the particular $\tau_{0}$ value is perhaps not that
informative, but the comparative values for different pulsators in the
instability strip clearly will provide physical insights.  In
particular, the correlation between the increasing values of
$\tau_{0}$ with the decreasing effective temperature estimates for
different pulsators looks to be very useful.  We discuss this feature
in the next section.

\section[]{Discussion}\label{sec:discussion}

The nonlinear lightcurve fitting procedure has been applied to a
number of stars in both the DAV and DBV instability strips, and the
correlations between the derived values of $\tau_{0}$ and both the
effective temperatures and mean pulsation periods for the different
stars have been investigated \citep{montgomery10,provencal12a}.

The most interesting correlation is the observed increase in the mean
convective response time $\tau_{0}$ for a given pulsator versus the
decrease in the spectroscopically measured effective temperature.  The
physical processes underlying this effect appear to be well
understood.  When a star enters the hot blue edge of the relevant
instabilty strip, the convection zone is quite thin and it therefore
has only a small impact on the flux variations passing through its
layers.  As the star cools, this convection zone deepens and becomes
more massive, and its effect is to distort the sinudoidal flux
variations by introducing increasing nonlinearities; these are
modelled in terms of larger $\tau_{0}$ values.

\begin{table}
\label{table:dav}
\caption{The measured convection zone parameter $\tau_{0}$ versus
  effective temperatures for a selection of DAV pulsators.  The values
  have been taken from \citet{provencal12a} with some rounding so as
  to reflect the various uncertainties.}
\centering
\begin{tabular}{lrl}
\hline
\multicolumn{1}{c}{DAV White Dwarf} & 
\multicolumn{1}{c}{$\tau_{0}$ [s]}   &
\multicolumn{1}{c}{T$_{\rm eff}$ [K]} \\ 
\hline
G\,117$-$B15A     &  30 $\pm$ 10   & $12\,000 \pm 200$ \\
EC\,14012$-$1446  & 100 $\pm$ 10   & $11\,770 \pm 25$  \\
G\,29$-$38        & 190 $\pm$ 20   & $11\,690 \pm 120$ \\
WDJ\,1524$-$0030  & 160 $\pm$ 35   & $11\,660 \pm 180$ \\
GD\,154           & 1170 $\pm$ 200 & $11\,270 \pm 170$ \\
\hline
\end{tabular}
\end{table}

The DAV pulsators provide a reasonable empirical test of this model as
the instability strip is now well populated and the spectroscopic
effective temperature estimates are sufficiently accurate to identify
a range of values within it.

In the case of the DBV instability strip there are far fewer known
pulsators, but even more importantly the spectroscopic determination
of the T$_{\rm eff}$ values are relatively uncertain, largely due to
the lack of sensitivity of the He I lines to changes in temperature in
the vicinity of 25\,kK \citep[e.g.][]{bergeron11}.  In addition, small
changes in an unknown H contamination in an otherwise pure He
atmosphere leads to different inferred T$_{\rm eff}$ values.

The DBV, EC\,04207 being studied in this paper provides an example.
The two published spectroscopic measurements of T$_{\rm eff}$ differ
by $\sim$2000\,K -- 25,000\,K \citep{koester01} and 27,300\,K
\citep{voss07}.

\begin{table}
\label{table:dbv}
\caption{The measured convection zone parameters $\tau_{0}$ for five
  DBV white dwarfs in order of the increasing effective temperatures
  as gauged from the decreasing values of $\tau_{0}$.  The $\tau_{0}$
  value for PG\,1351 is taken from \citet{montgomery05} and the values
  for GD\,358 and PG\,1654 are from \citet{montgomery10}.  Note that
  the second entry for GD\,358 with the much smaller $\tau_{0}$ = 42\,s
  value corresponds to a surprising temporary change in the lightcurve
  behaviour of this well studied pulsator, which has been interpreted
  by \citet{montgomery10} as a temporary increase in the star's
  T$_{\rm eff}$ of $\sim2000$\,K.}
\centering
\begin{tabular}{lr}
\hline
\multicolumn{1}{c}{DBV White Dwarf} & 
\multicolumn{1}{c}{$\tau_{0}$ [s]}  \\
\hline
GD\,358           & 570 $\pm$ 6 \\
EC\,04207$-$4748  & 148 $\pm$ 5 \\
PG\,1654$+$160    & 117 $\pm$ 5 \\
PG\,1351$+$489      &  88 $\pm$ 8 \\
GD\,358*          &  42 $\pm$ 2 \\
EC\,20058$-$5234  &  40 $\pm$ 5 \\
\hline
\end{tabular}
\end{table}

It would appear to be fruitful to use the $\tau_{0}$ values
established from the lightcurve fitting process to identify the
various pulsators within the DBV instability strip, and then use them
to act as a check on the spectroscopic temperatures or even
provide an alternative indirect method of establishing their relative
temperatures, and therefore their positions within the instability
strip.

\cite{montgomery10} cite $\tau_{0}$ values extracted from the light
curves of the pulsators GD\,358, PG\,1351+489 and PG\,1654+160.  We
also report here premiminary results for lightcurve analysis of the
DBV EC\,20058$-$5234 \citep{sullivan07}.  The lightcurve of this hot
white dwarf pulsator exhibits no obvious departures from a sinusoidal
form, although there is evidence in the DFTs of small contributions
from harmonics and combination frequencies.  Using the five dominant
pulsation modes (333, 281, 257 333 and 195 [s]) in this star
\citep{sullivan08a}, we have extracted a value of $\tau_{0} \sim
40$\,s for EC\,20058 from lightcurve fitting.

In Table \ref{table:dbv} we have listed these five DBVs in terms of
their decreasing numerical values for the deduced average convective
response times $\tau_{0}$.  This table clearly suggests that EC\,20058
with the smallest $\tau_{0}$ value is closest to the blue edge of the
DBV instability strip and therefore has the highest T$_{\rm eff}$,
while the class prototype GD\,358 (ie V777) in its normal or quiescent
state \citep{montgomery10} with the largest $\tau_{0}$ value is
closest to the red edge and has the lowest T$_{\rm eff}$.

\begin{figure}
\centering
\includegraphics[width=7.7cm, angle=-90]{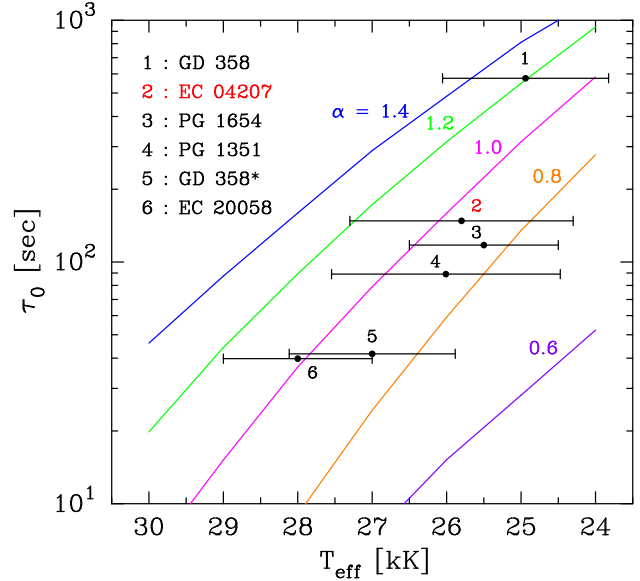}
\caption{Model convective response times $\tau_{0}$ as a function of
  T$_{\rm eff}$ for different values of the $\alpha$ parameter using
  the ML2 convection model.  Note that the measured $\tau_{0}$ value
  for EC\,04027 is broadly consistent with $0.8 \le \alpha \le 1.2$
  using the range of possible T$_{\rm eff}$ values we have adopted for
  this star. (A colour version of this figure is available in the
    online journal.)}
\label{fig:mlalpha}
\end{figure}

In Fig.\ \ref{fig:mlalpha} we have presented expected values for
$\tau_{0}$ determined using the ML2 convection model
\citep{bohm71,provencal12a}, and plotted these as a function of
effective temperature in the vicinity of the DBV instability strip.
As is depicted in the figure, we computed model $\tau_{0}$ values for
five different values of the ML2 $\alpha$ parameter, which corresponds
to the mixing length relative to the atmosphere pressure scale height.
The various measured $\tau_{0}$ values for the five DBVs have been
added to the plot along with spectroscopically estimated effective
temperatures for each star.  The T$_{\rm eff}$ estimates we have used
are from the following sources: (1) GD\,358 and PG\,1351
\citep{bergeron11}, (2) EC\,04207 and PG\,1654
\citep{voss07,koester01} and (3) EC\,20058 (D.\ Koester, private
communication).  It is important to note that the error bars depicted
in Fig.\ \ref{fig:mlalpha} are not really formal uncertainlty
estimates, but are in reality indicative intervals for the
spectroscopically estimated T$_{\rm eff}$ values.

Note also that \cite{bergeron11} also lists a value for PG\,1654 of
T$_{\rm eff} = 29\,410 \pm 1613$.  This appears unreasonably high to
us, so either this is a mistake or given the implied range of more
than 3\,000\,K it is very suggestive of the real difficulty of
establishing a spectroscopic temperature in the DBV temperature range.
In Fig.\ \ref{fig:mlalpha} we have used the lower T$_{\rm eff}$
published in \cite{voss07} along with an appropriate uncertainty
estimate.

In a future publication we aim to relate in more detail the measured
$\tau_{0}$ values to the spectroscopic measurements of the
temperatures.

\section*{Acknowledgments}
We thank the Marsden Fund of NZ for providing financial support for
this research and the University of Canterbury for the allocation of
telescope time for the project.  We also thank an anonymous referee
for constructive comments that led to an improvement in the paper.

\bibliographystyle{mn2e}
\bibliography{sullivan12}

\begin{thebibliography}{}

\bibitem[\protect\citeauthoryear{{Althaus}, {C{\'o}rsico}, {Isern} \&
  {Garc{\'{\i}}a-Berro}}{{Althaus} et~al.}{2010}]{althaus10}
{Althaus} L.~G.,  {C{\'o}rsico} A.~H.,  {Isern} J.,    {Garc{\'{\i}}a-Berro}
  E.,  2010, A\&ARv, 18, 471

\bibitem[\protect\citeauthoryear{{Bergeron}, {Wesemael}, {Dufour}, {Beauchamp},
  {Hunter}, {Saffer}, {Gianninas}, {Ruiz}, {Limoges}, {Dufour}, {Fontaine} \&
  {Liebert}}{{Bergeron} et~al.}{2011}]{bergeron11}
{Bergeron} P.,  {Wesemael} F.,  {Dufour} P.,  {Beauchamp} A.,  {Hunter} C.,
  {Saffer} R.~A.,  {Gianninas} A.,  {Ruiz} M.~T.,  {Limoges} M.-M.,  {Dufour}
  P.,  {Fontaine} G.,    {Liebert} J.,  2011, ApJ, 737, 28

\bibitem[\protect\citeauthoryear{{B\"{o}hm} \& {Cassinelli}}{{B\"{o}hm} \&
  {Cassinelli}}{1971}]{bohm71}
{B\"{o}hm} K.~H.,  {Cassinelli} J.,  1971, aa2, 12, 21B

\bibitem[\protect\citeauthoryear{Brickhill}{Brickhill}{1992}]{brickhill92a}
Brickhill A.~J.,  1992, MNRAS, 259, 519

\bibitem[\protect\citeauthoryear{{Fontaine} \& {Brassard}}{{Fontaine} \&
  {Brassard}}{2008}]{fontaine08}
{Fontaine} G.,  {Brassard} P.,  2008, PASP, 120, 1043

\bibitem[\protect\citeauthoryear{{Goldreich} \& {Wu}}{{Goldreich} \&
  {Wu}}{1999}]{goldreich99a}
{Goldreich} P.,  {Wu} Y.,  1999, ApJ, 511, 904

\bibitem[\protect\citeauthoryear{{Hermes}, {Montgomery}, {Winget}, {Brown},
  {Kilic} \& {Kenyon}}{{Hermes} et~al.}{2012}]{hermes12}
{Hermes} J.~J.,  {Montgomery} M.~H.,  {Winget} D.~E.,  {Brown} W.~R.,  {Kilic}
  M.,    {Kenyon} S.~J.,  2012, apj2, 750, L28

\bibitem[\protect\citeauthoryear{Kilkenny, O'Donoghue, Crause, Hambly \&
  MacGillivray}{Kilkenny et~al.}{2009}]{kilkenny09}
Kilkenny D.,  O'Donoghue D.,  Crause L.~A.,  Hambly N.,    MacGillivray H.,
  2009, MNRAS, 397, 453

\bibitem[\protect\citeauthoryear{{Koester}, {Napiwotzki}, {Christlieb},
  {Drechsel}, {Hagen}, {Heber}, {Homeier}, {Karl}, {Leibundgut}, {Moehler},
  {Nelemans}, {Pauli}, {Reimers}, {Renzini} \& {Yungelson}}{{Koester}
  et~al.}{2001}]{koester01}
{Koester} D.,  {Napiwotzki} R.,  {Christlieb} N.,  {Drechsel} H.,  {Hagen}
  H.-J.,  {Heber} U.,  {Homeier} D.,  {Karl} C.,  {Leibundgut} B.,  {Moehler}
  S.,  {Nelemans} G.,  {Pauli} E.-M.,  {Reimers} D.,  {Renzini} A.,
  {Yungelson} L.,  2001, A\&A, 378, 556

\bibitem[\protect\citeauthoryear{Landolt}{Landolt}{1968}]{landolt68}
Landolt A.~U.,  1968, ApJ, 153, 151

\bibitem[\protect\citeauthoryear{{Montgomery}}{{Montgomery}}{2005}]{montgomery05}
{Montgomery} M.~H.,  2005, ApJ, 633, 1142

\bibitem[\protect\citeauthoryear{{Montgomery}, {Provencal}, {Kanaan},
  {Mukadam}, {Thompson}, {Dalessio}, {Shipman}, {Winget}, {Kepler} \&
  {Koester}}{{Montgomery} et~al.}{2010}]{montgomery10}
{Montgomery} M.~H.,  {Provencal} J.~L.,  {Kanaan} A.,  {Mukadam} A.~S.,
  {Thompson} S.~E.,  {Dalessio} J.,  {Shipman} H.~L.,  {Winget} D.~E.,
  {Kepler} S.~O.,    {Koester} D.,  2010, ApJ, 716, 84

\bibitem[\protect\citeauthoryear{Mukadam, Mullally, Nather, Winget, von Hippel,
  Kleinman, Nitta, Krzesinski et~al.,}{Mukadam et~al.}{2004}]{mukadam04a}
Mukadam A.~S.,  Mullally F.,  Nather R.~E.,  Winget D.~E.,  von Hippel T.,
  Kleinman S.~J.,  Nitta A.,  Krzesinski J.,    et~al., 2004, ApJ, 607, 982

\bibitem[\protect\citeauthoryear{Mullally, Thompson, Castanheira, Winget,
  Kepler, Eisenstein, Kleinman \& Nitta}{Mullally et~al.}{2006}]{mullally06}
Mullally S.~E.,  Thompson S.~E.,  Castanheira B.~G.,  Winget D.~E.,  Kepler
  S.~O.,  Eisenstein D.~J.,  Kleinman S.~J.,    Nitta A.,  2006, ApJ, 640, 956

\bibitem[\protect\citeauthoryear{Nather \& Mukadam}{Nather \&
  Mukadam}{2004}]{nather04}
Nather R.~E.,  Mukadam A.~S.,  2004, ApJ, 605, 846

\bibitem[\protect\citeauthoryear{Nather, Winget, Clemens, Hansen \&
  Hine}{Nather et~al.}{1990}]{nather90}
Nather R.~E.,  Winget D.~E.,  Clemens J.~C.,  Hansen C.~J.,    Hine B.~P.,
  1990, ApJ, 361, 309

\bibitem[\protect\citeauthoryear{Nitta, Kleinman, Krzesinski, Kepler, Metcalfe,
  Mukadam, Mullally, Nather, Sullivan, Thompson \& Winget}{Nitta
  et~al.}{2009}]{nitta09}
Nitta A.,  Kleinman S.~J.,  Krzesinski J.,  Kepler S.~O.,  Metcalfe T.~S.,
  Mukadam A.~S.,  Mullally F.,  Nather R.~E.,  Sullivan D.~J.,  Thompson S.~E.,
     Winget D.~E.,  2009, ApJ, 690, 560

\bibitem[\protect\citeauthoryear{{{\O}stensen}, {Bloemen}, {Vu{\v c}kovi{\'c}},
  {Aerts}, {Oreiro}, {Kinemuchi}, {Still} \& {Koester}}{{{\O}stensen}
  et~al.}{2011}]{ostensen11}
{{\O}stensen} R.~H.,  {Bloemen} S.,  {Vu{\v c}kovi{\'c}} M.,  {Aerts} C.,
  {Oreiro} R.,  {Kinemuchi} K.,  {Still} M.,    {Koester} D.,  2011, apj2, 736,
  L39

\bibitem[\protect\citeauthoryear{{Provencal} et~al.,}{{Provencal}
  et~al.}{2012}]{provencal12a}
{Provencal} J.~L.,  et~al., 2012, ApJ, 751, 91

\bibitem[\protect\citeauthoryear{Scargle}{Scargle}{1982}]{Scargle82}
Scargle J.~D.,  1982, ApJ, 263, 835

\bibitem[\protect\citeauthoryear{Stobie, Kilkenny, O'Donoghue, Chen, Koen,
  Morgan, Barrow, Buckley D, Cannon, Cass et~al.,}{Stobie
  et~al.}{1997}]{stobie97}
Stobie R.~S.,  Kilkenny D.,  O'Donoghue D.,  Chen A.,  Koen C.,  Morgan D.~H.,
  Barrow J.,  Buckley D A.~H.,  Cannon R.~D.,  Cass C. J.~P.,    et~al., 1997,
  MNRAS, 287, 848

\bibitem[\protect\citeauthoryear{Sullivan}{Sullivan}{2000}]{sullivan00a}
Sullivan D.~J.,  2000, Baltic Astronomy, 9, 425

\bibitem[\protect\citeauthoryear{{Sullivan} et~al.,}{{Sullivan}
  et~al.}{2008}]{sullivan08a}
{Sullivan} D.~J.,  et~al., 2008, MNRAS, 387, 137

\bibitem[\protect\citeauthoryear{Sullivan, Metcalfe, O'Donoghue, Winget
  et~al.,}{Sullivan et~al.}{2007}]{sullivan07}
Sullivan D.~J.,  Metcalfe T.,  O'Donoghue D.,  Winget D.~E.,    et~al., 2007,
  in {Napiwotzki} R.,  {Burleigh} M.,  eds, 15$^{th}$ European Workshop on
  White Dwarfs Vol.~372 of ASP Conf..
Astron. Soc. Pac., San Francisco, p.~629

\bibitem[\protect\citeauthoryear{{Voss}, {Koester}, {Napiwotzki}, {Christlieb}
  \& {Reimers}}{{Voss} et~al.}{2007}]{voss07}
{Voss} B.,  {Koester} D.,  {Napiwotzki} R.,  {Christlieb} N.,    {Reimers} D.,
  2007, A\&A, 470, 1079

\bibitem[\protect\citeauthoryear{{Winget} \& {Kepler}}{{Winget} \&
  {Kepler}}{2008}]{winget08}
{Winget} D.~E.,  {Kepler} S.~O.,  2008, Ann.Rev.Astron.Ap., 46, 157

\bibitem[\protect\citeauthoryear{Winget, Robinson, Nather \& Fontaine}{Winget
  et~al.}{1982}]{winget82}
Winget D.~E.,  Robinson E.~L.,  Nather R.~E.,    Fontaine G.,  1982, ApJ, 262,
  L11

\end{thebibliography}

\label{lastpage}

\end{document}